\documentclass[english,prb,aps,twocolumn,10pt]{revtex4-2}

\usepackage{amsmath}
\usepackage{amssymb}
\usepackage{graphicx} 
\usepackage{babel}
\usepackage{bm} 
\usepackage{color} 
\usepackage{verbatim}
\usepackage[colorlinks=true, pdfstartview=FitV, linkcolor=blue, citecolor=blue, urlcolor=blue]{hyperref} 
\usepackage{lineno}

\newcommand{\captionlinespread}{1.10}

\def\Cr				{{Cr$^{3+}$ \/}}
\def\CrO			{{Cr$_2$O$_3$ \/}}
\def\CrOa			{{Cr$_2$O$_3$}}
\def\O				{{O$^{2-}$ \/}}

\newcommand{\mr}[1]{\mathrm{#1}}
\newcommand{\unit}[1]{\,\mathrm{#1}}
\newcommand{\um}{\,\mu{\rm m}}

\newcommand{\muB}{\mu_{\rm B}}

\newcommand{\rms}{\text{-rms}}

\newcommand{\Dw}{\Delta}
\newcommand{\Lp}{L^{+}}
\newcommand{\Lm}{L^{-}}
\newcommand{\Ms}{\sigma_z^0}

\newcommand{\TN}{T_\mr{N}}
\newcommand{\xo}{x_0}
\newcommand{\Kp}{K_\mr{ip}}
\newcommand{\psiip}{\psi_\mr{ip}}

\newcommand{\sigx}{\sigma_x}
\newcommand{\sigy}{\sigma_y}
\newcommand{\sigz}{\sigma_z}

\newcommand{\Bx}{B_x}

\newcommand{\BNV}{B_\mr{NV}}

\begin{document}

\title{Co-existence of Bloch and N\'eel walls in a collinear antiferromagnet}
%
\author{M. S. W\"ornle$^{1,2,\dagger}$}
\author{P. Welter$^{1,\dagger}$}
\author{M. Giraldo$^{2,\dagger}$}
\author{T. Lottermoser$^2$}
\author{M. Fiebig$^2$\footnote{Email:manfred.fiebig@mat.ethz.ch}}
\author{P. Gambardella$^2$\footnote{Email:pietro.gambardella@mat.ethz.ch}}
\author{C. L. Degen$^1$\footnote{Email:degenc@ethz.ch}}
\email{degenc@ethz.ch}
\email{$^\dagger$These authors contributed equally to this work}
\affiliation{$^1$Department of Physics, ETH Zurich, Switzerland}
\affiliation{$^2$Department of Materials, ETH Zurich, Switzerland}

\date{\today}


\begin{abstract}
We resolve the domain-wall structure of the model antiferromagnet $\text{Cr}_2\text{O}_3$ using nanoscale scanning diamond magnetometry and second-harmonic-generation microscopy. We find that the 180$^\circ$ domain walls are predominantly Bloch-like, and can co-exist with N\'eel walls in crystals with significant in-plane anisotropy. In the latter case, N\'eel walls that run perpendicular to a magnetic easy axis acquire a well-defined chirality.  We further report quantitative measurement of the domain-wall width and surface magnetization.  Our results provide fundamental input and an experimental methodology for the understanding of domain walls in pure, intrinsic antiferromagnets, which is relevant to achieve electrical control of domain-wall motion in antiferromagnetic compounds.
\end{abstract}
%


\maketitle


\section{Introduction}

One of the great unknowns of antiferromagnetism is the domain wall that separates regions with different orientation of the magnetic order parameter. The domain-wall structure influences the thermal stability \cite{Shpyrko2007}, exchange bias \cite{Radu2008}, and magnetoresistance \cite{Jaramillo2007,Wornle2019} of antiferromagnets. Furthermore, the type of domain wall, Bloch or N\'{e}el, determines their response to current-induced spin torques \cite{Hals2011,Gomonay2016,Shiino2016,Baldrati2019}, which is of relevance for emerging applications in spintronics of both intrinsic \cite{Jungwirth2016,Baltz2018,Manchon2019} and synthetic \cite{Duine2018,Yang2015,Luo2019,Hrabec2020} antiferromagnets.

Unlike for ferromagnets \cite{Hubert2008}, the internal structure of domain walls in antiferromagnets is not generally known.  Whereas the antiferromagnetic domain pattern has been imaged for a number of materials including intrinsic antiferromagnets, multiferroics and magnetically-coupled thin films \cite{Geng2014,Cheong2020,Hellwig2003}, these studies generally do not consider the detailed internal domain wall structure.  Exceptions include a few systems where antiferromagnetic order is accompanied by strain \cite{Weber2003} or defects \cite{Ravlic2003}, monolayer-thin films \cite{Bode2006}, and synthetic antiferromagnets \cite{Yang2015}.  By contrast, to the best of our knowledge, no studies for bulk, intrinsic antiferromagnets have been reported.  Theoretical analysis suggests that, in the absence of in-plane magnetic anisotropy or a Dzyaloshinskii-Moriya interaction (DMI), no preference is expressed for either Bloch or N\'{e}el walls \cite{Malozemoff2016,Gyorgy1968,Papanicolaou1995,Mitsumata2011,Chen2019}.  The limited experimental knowledge about antiferromagnetic domain walls is due to a lack of techniques capable of spatially resolving the internal wall structure.

In this work, we use nanoscale scanning diamond magnetometry (NSDM) to investigate the spin structure of the pure intrinsic antiferromagnet \CrOa.  NSDM microscopy is an emerging quantum technique for the imaging of weak magnetic fields with nanometer spatial resolution (Fig. \ref{fig1}), with remarkable progress on antiferromagnets \cite{Kosub2017,Appel2019,Wornle2019}, multiferroics \cite{Gross2017}, and helimagnets \cite{Dussaux2016}.  Here, we extend NSDM to the imaging of antiferromagnetic $180^\circ$ domain-wall structures.  We obtain quantitative information about the domain-wall width, chirality and surface magnetization, and connect it to a model of interplaying demagnetizing and anisotropy energies.  We find that both Bloch and N\'eel walls can be present.  Our work extends the knowledge about antiferromagnetic domain wall structure to the most basic class of intrinsic, bulk antiferromagnets.
%



\section{One-dimensional model of antiferromagnetic domain walls}
\label{1Dmodel}

\begin{figure}[t]
    \centering
    \includegraphics[width=0.98\columnwidth]{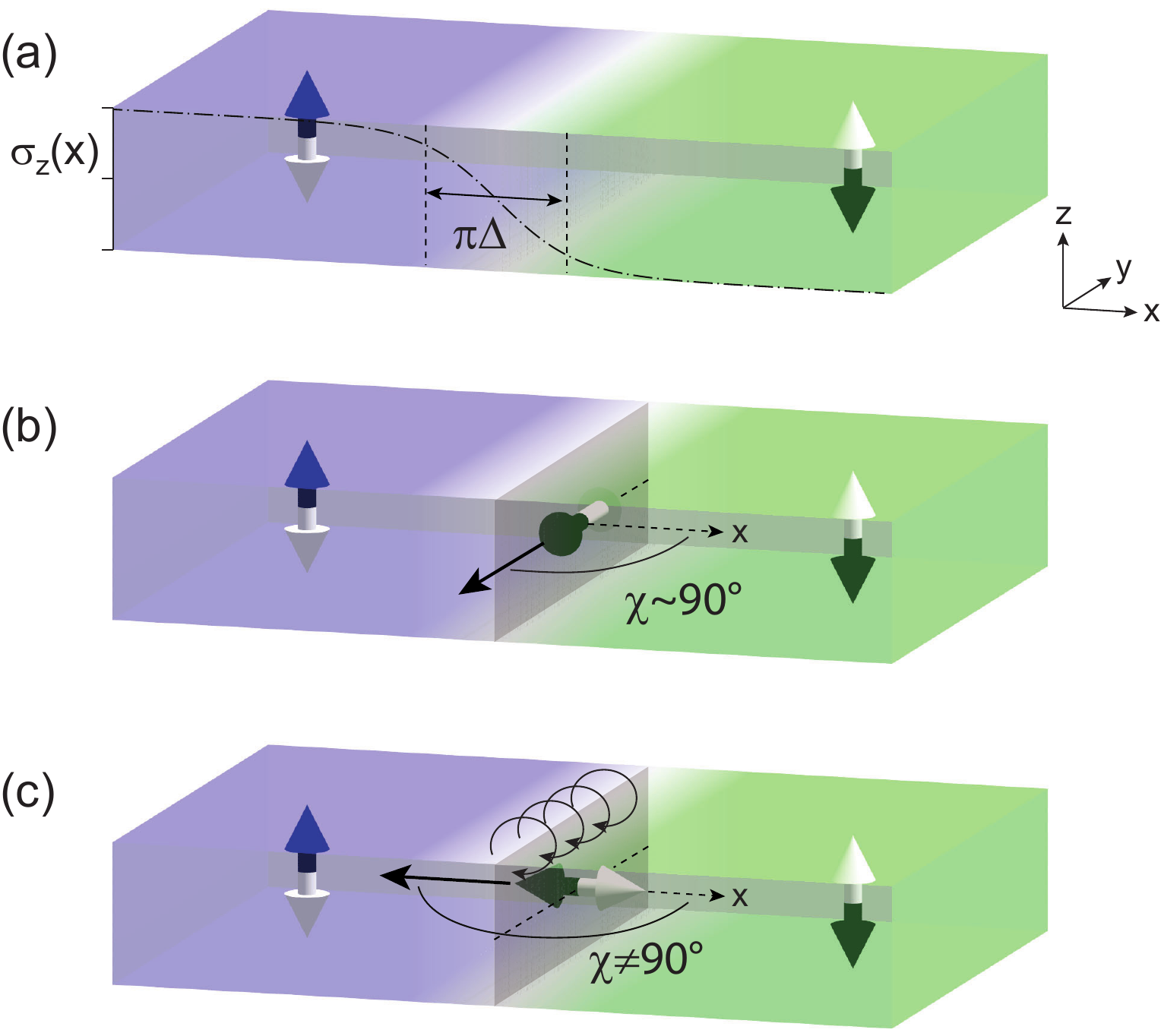}
    \caption{
		\linespread{\captionlinespread}\selectfont{}
{One-dimensional model for an antiferromagnetic $180^\circ$ domain wall.}
{(a)} Sketch of the domain wall separating regions (purple, green) of opposite order parameter.  The dash-dotted line describes the wall profile as given by Eq.~(\ref{eq:sigwallz}).
{(b)} The presence of a residual demagnetizing field favors the formation of Bloch walls ($\chi = \pm\pi/2$). {(c)} For sufficiently large in-plane anisotropy in a direction orthogonal to the wall, the formation of a N\'eel wall ($\chi = 0,\,\pi$) or mixed N\'eel-Bloch wall is favored. Curled arrows indicate the demagnetizing field arising from moments crossing the domain wall perpendicularly.
}
\label{fig0}
\end{figure}

\begin{figure*}[t]
    \centering
    \includegraphics[width=0.74\textwidth]{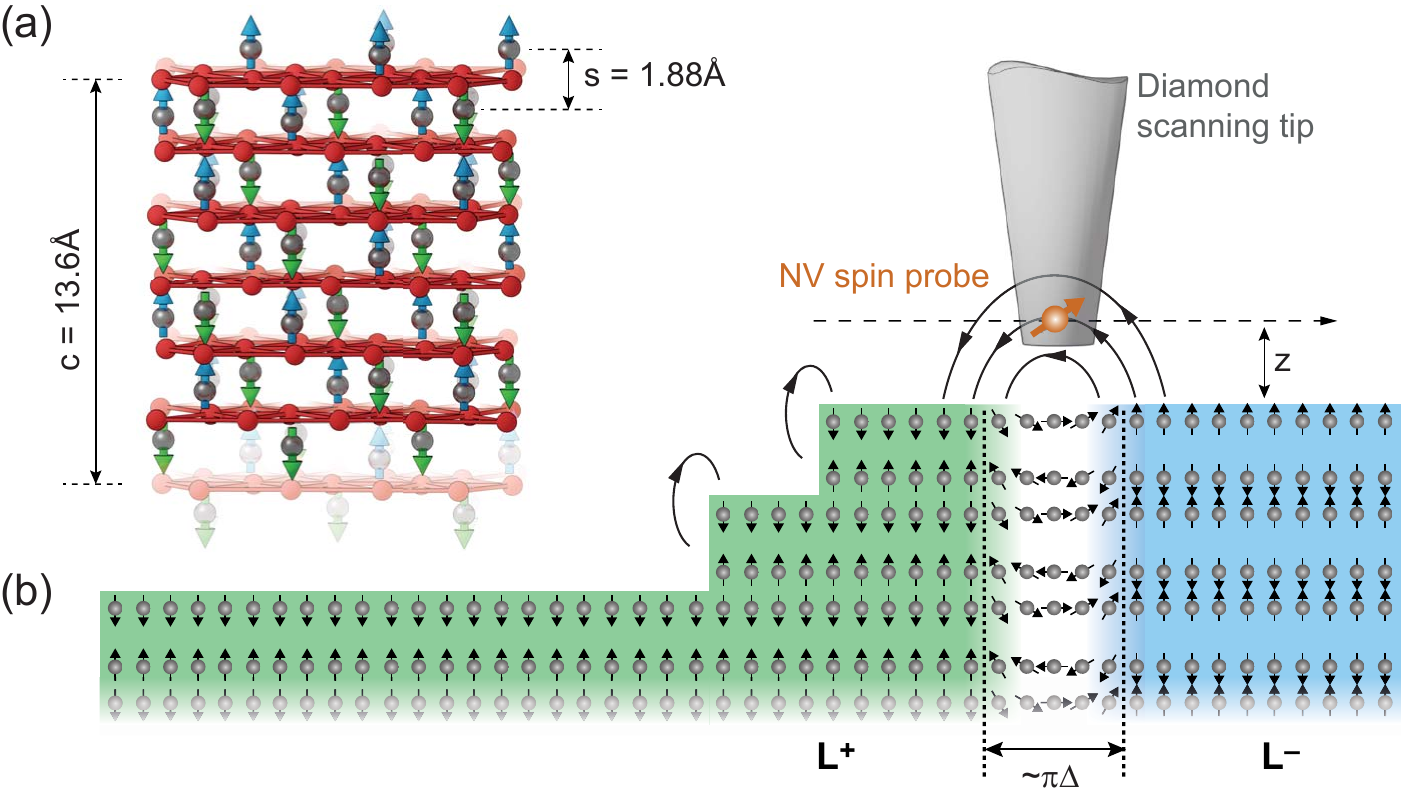}
    \caption{
		\linespread{\captionlinespread}\selectfont{}
{\CrO crystal structure and experimental arrangement.}
{(a)} Side view of the hexagonal unit cell.  Blue and green arrows symbolize \Cr moments of opposite magnetic polarization, red atoms are O$^{2-}$ ions.
{(b)} Lateral cut through the $c$-oriented \CrO sample surface.  Strong magnetic stray fields (black field lines) are expected at antiferromagnetic domain walls and weak fields at monolayer topographic steps.  Blue and green shading indicate regions of opposite order parameter $\Lp$ and $\Lm$, defined by the orientation (up or down) of the topmost \Cr atom in the unit cell \cite{Fiebig1996}.  The regions are separated by a domain wall (white) of approximate width $\pi\Dw$. The diamond scanning tip and NV center are shown in gray and orange, respectively.
%
    }
    \label{fig1}
\end{figure*}

In order to motivate and explain our experimental observations, we briefly review the conventional model for static one-dimensional domain walls \cite{Malozemoff2016,Gyorgy1968} and extend it to collinear antiferromagnets.  We consider a $180^{\circ}$ domain wall, as it occurs in, for example, \CrOa, $\alpha$-Fe$_2$O$_3$, or CuMnAs.  The domain wall separates two regions with magnetic order parameter pointing up ($x<0$) and down  ($x>0$), as shown schematically in Fig. \ref{fig0}. The key parameters of such a domain wall are the wall width $\Dw$ and the twist angle $\chi$ between the wall magnetization and the $x$-axis. Using this notation, Bloch walls correspond to $\chi = \pm \pi/2$ and N\'eel walls to $\chi = 0$ ($\pi$) for walls with right (left) chirality.   $\Dw$ and $\chi$ are determined by the interplay between exchange and anisotropy energies, and further modified by the demagnetizing field and DMI, if present.  Considering only the first two contributions, the local domain wall energy density in the continuum limit reads:
\begin{align}
e = A \left[\left(\frac{\partial\theta}{\partial x}\right)^2
	+ \sin^2\theta \left(\frac{\partial\phi}{\partial x}\right)^2\right] + K\sin^2\theta,
\end{align}
where $A$ is the exchange stiffness, $K$ is the uniaxial anisotropy constant, and $(\theta,\phi)$ are the polar coordinates of the order parameter defined relative to the $z$-axis shown in Fig. \ref{fig0}.
The static equilibrium solution that minimizes the total wall energy and satisfies the boundary conditions $\theta\left(\pm\infty\right)=\left(0,\pi\right)$ is given by:
\begin{subequations} 
\label{eq:sigwallpolar}
\begin{align}
\phi\left(x\right) &= \chi = \mathrm{const.} \\
\theta\left(x\right) &= \pm 2 \arctan\left[\exp\left(x/\Dw\right)\right],
\end{align}
\end{subequations}
where $\Dw = \Dw_0 := \sqrt{A/K}$ is the domain wall width \cite{Malozemoff2016,Papanicolaou1995,Mitsumata2011}.
The Cartesian coordinates of the order parameter can be easily derived from Eq.~(\ref{eq:sigwallpolar}). For a layered antiferromagnet like \CrOa, it is convenient to express $\theta(x)$ and $\phi(x)$ in terms of the intrinsic surface magnetization $\vec\sigma(x)$:
\begin{subequations} 
\label{eq:sigwall}
\begin{align}
\sigx(x) &= \Ms \left[ \cosh\left(\frac{x-\xo}\Dw\right) \right]^{-1} \cos\chi \label{eq:sigwallx} \ , \\
\sigy(x) &= \Ms \left[ \cosh\left(\frac{x-\xo}\Dw\right) \right]^{-1} \sin\chi \label{eq:sigwally} \ , \\
\sigz(x) &= \Ms \tanh\left(\frac{x-\xo}\Dw\right) \label{eq:sigwallz} \ ,
\end{align}
\end{subequations}
where we assume a domain wall centered at $x=\xo$. The profile of $\sigz(x)$ is shown by a dash-dotted curve in Fig. \ref{fig0}(a).  The total energy per unit area of the wall $\epsilon_0$ is given by $\epsilon_0 = 4\sqrt{AK}$.

We stress that, up to this point, the twist angle $\chi$ is arbitrary and independent of $x$.  In other words, for an antiferromagnetic system governed solely by exchange and anisotropy energies, N\'eel and Bloch domain walls (or any combination of the two) are degenerate in energy.

To understand the preference for one type of wall over the other, we next consider the effects of a demagnetizing field $\mu_0 M$ and of an in-plane magnetic anisotropy energy density $\Kp$. We note that, although the volume magnetization $M$ in an antiferromagnet is zero, a finite demagnetizing field still persists when $\phi \neq \pm \pi/2$ due to the net magnetic moment of the domain wall~\cite{Papanicolaou1995,Tveten2016}. Keeping in mind these additional contributions, the domain wall energy per unit area changes to~\cite{Malozemoff2016}:
\begin{equation}
\epsilon = \epsilon_0 + \mu_0M^2 \Dw_0\cos^2\chi + 2\Kp\Dw_0\sin^2\left(\chi-\psiip\right),
\label{eq:dwenergy2}
\end{equation}
where the in-plane easy axis is defined by the angle $\psiip$ relative to the $x$-axis.  According to Eq.~(\ref{eq:dwenergy2}), the residual demagnetizing field favors the formation of Bloch walls over N\'eel walls (Fig. \ref{fig0}(b)). On the other hand, the in-plane anisotropy forces the magnetic moments to cant along the in-plane easy axis, leading to a competition between demagnetizing and anisotropy energies. For a sufficiently large in-plane anisotropy $2\Kp>\mu_0 M^2$ favoring the $x$-direction ($\psiip = 0$), we thus expect a N\'eel-type or a mixed N\'eel/Bloch-type domain wall (Fig. \ref{fig0}(c)).  

The presence of a residual demagnetizing field and of in-plane anisotropy also leads to a modification of the domain wall width:
\begin{equation}
\Dw = \Dw_0\left[1-\frac{\mu_0M^2}{4K}\cos^2\chi-\frac{\Kp}{2K}\sin^2\left(\chi-\psiip\right)\right].
\label{eq:dwMagnetostatics}
\end{equation}
In particular, Eq.~(\ref{eq:dwMagnetostatics}) predicts that the width of a N\'eel wall is reduced with respect to a Bloch wall, with a ratio given by:
\begin{align}
\frac{\Dw_{\text{N\'eel}}}{\Dw_{\text{Bloch}}}
	= 1-\frac{\mu_0M^2}{4K}
\label{eq:reductiondw}
\end{align}
where $M$ is approximately given by the magnetization of the polarized surface layer.  

In addition to these interactions, the DMI can further lead to a preference for N\'eel-type domain walls when the Dzyaloshinskii vector runs parallel to the domain wall direction (the $y$-axis, Fig.~\ref{fig0}) \cite{Malozemoff2016,Gyorgy1968,Heide2008}. Although the DMI is zero in bulk monodomain \CrO for symmetry reasons \cite{Dzyaloshinsky1958}, the formation of two magnetic domains with order parameter along $\pm z$ breaks the inversion symmetry between two adjacent spins along $x$. Such a symmetry breaking might allow for a finite DMI or higher-order chiral interactions to emerge in proximity of domain walls in \CrOa \cite{Moriya1960}. If such local chiral interactions are larger or comparable to the demagnetizing energy, the domain walls will be of either N\'eel or intermediate Bloch-N\'eel-type with a unique chirality.  The results presented in our study provide experimental support to this hypothesis.

%
%


\section{Materials and Methods}

We investigate the domain wall structure and chirality in the prototypical 180$^\circ$ antiferromagnet \CrOa.  \CrO is an antiferromagnetic insulator consisting of a hexagonal close packed array of \O anions with $2/3$ of the octahedral holes occupied by \Cr \cite{Fiebig1996} (Fig.~\ref{fig1}a).  Below $\TN=307.6\unit{K}$, \CrO forms an antiferromagnetically ordered phase, where the \Cr ions organize in alternating layers of opposite magnetic polarization (green and blue spins in Fig.~\ref{fig1}a).  Because of its fundamental role in antiferromagnetism, \CrO has served as a model system for uniaxial antiferromagnetic order \cite{Dzyaloshinsky1958,Rado1961,Brockhouse1953}, magnetoelectric coupling \cite{Astrov1961,Martin1966,Mostovoy2010}, and electrically controlled exchange bias \cite{Borisov2005,He2010}.  More recently, \CrO has attracted attention as a candidate material for antiferromagnetic magnetoelectric random access memories \cite{Kosub2017}, spin colossal magnetoresistance \cite{Qiu2018}, and as generator of sub-THz spin currents \cite{Li2020}.  Although the domain wall plays a critical role in many of these phenomena, the spin structure is unknown beyond initial theoretical work \cite{Kota2016}, presenting an important experimental test case.


We investigate the \CrO domain texture of three bulk single crystals.  Samples A and B are grown by the Verneuil method and polished to a surface roughness of $1-3\unit{nm\rms}$.  Sample C is a flux-grown platelet with an as-grown surface.  In a previous study \cite{Pisarev1997}, we found that the spin flop transition -- normally requiring a magnetic field of $5.8\unit{T}$ (Ref. \onlinecite{Fiebig1996}) -- occurs spontaneously at $150\unit{K}$ in sample C, pointing to an unusually strong in-plane anisotropy.  In addition, this sample has a lower N\'eel temperature ($\TN = 304.6\unit{K}$), probably due to strain or oxygen deficiency.
We create antiferromagnetic domains by repeatedly cooling samples through the transition temperature $\TN$ using magnetoelectric poling \cite{Krichevtsov1988} or until a multi-domain state spontaneously forms.  Further details about the samples are given in Ref. \onlinecite{supplementary}.

We use a combination of optical second-harmonic generation (SHG) microscopy and NSDM to locate the antiferromagnetic domains and measure the domain wall profile. SHG is a non-linear optical method capable of resolving the global 180$^\circ$ domain pattern, yet has a diffraction-limited spatial resolution and is not sensitive to the absolute sign of the order parameter \cite{Fiebig1995,supplementary}.
To map the stray field distribution, we scan a diamond tip with a nitrogen vacancy (NV) center (orange arrow) at constant height ($z=60-70\unit{nm}$) above the sample surface (Fig.~\ref{fig1}b).  The NV spin detects the component of the stray magnetic field $\BNV$ parallel to its internal anisotropy axis (here $55^\circ$ off the surface normal \cite{supplementary}).  The experiments are performed under ambient conditions at a temperature of 295~K.


\section{Experimental Results}

\subsection{Domain states}
\begin{figure}[t]
    \centering
    \includegraphics[width=1.00\columnwidth]{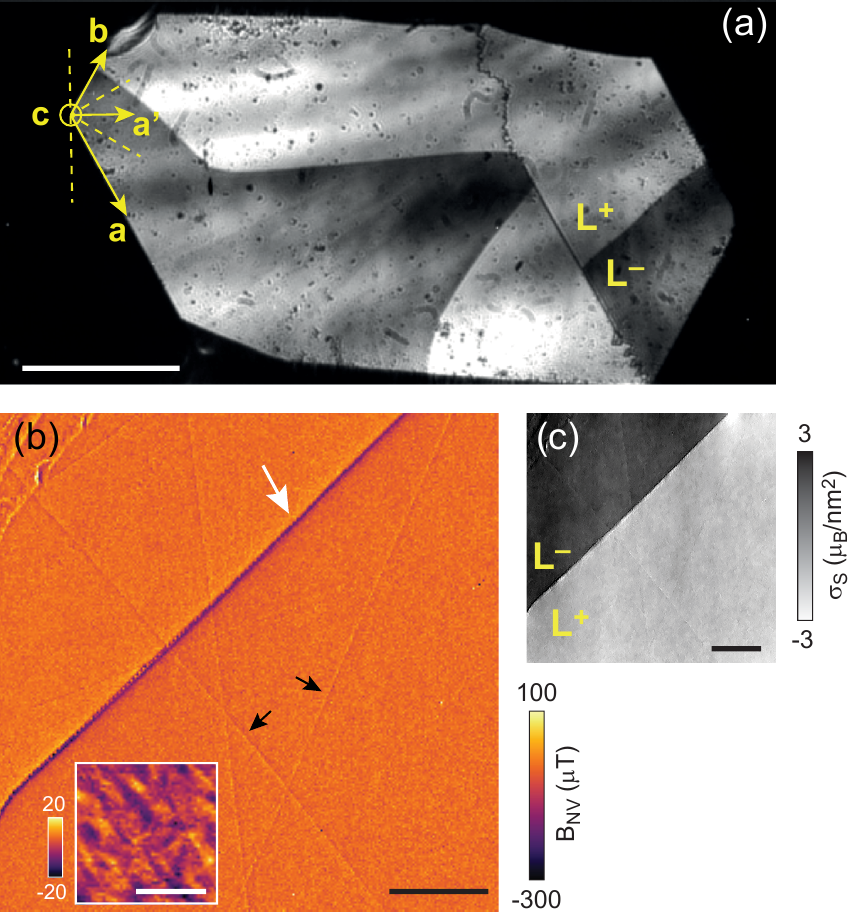}
    \caption{
		\linespread{\captionlinespread}\selectfont{}
{Antiferromagnetic domain pattern in $c$-oriented \CrOa.}
{(a)} SHG image revealing bright and dark domains of opposite order parameter in sample C; corresponding images for samples A and B are given in Figs. S1 and S2 in Ref. \onlinecite{supplementary}.  The order parameter ($\Lp$ and $\Lm$) is assigned based on the magnetization map in panel c.  The image is acquired with right-handed circularly-polarized illumination.  The two-fold axes \textit{a}, \textit{a'} and \textit{b} (yellow vectors), determined by X-ray crystallography, coincide with the in-plane magnetic easy axes of the spin flop phase.
Dashed lines indicate the in-plane magnetic hard axes. Scale bar, $500\unit{\um}$.
{(b)} Magnetometry image of the stray field above a domain wall (white arrow) in sample A.  Fainter features are due to surface topography, such as scratch marks from sample polishing (black arrows).  The inset shows a high-sensitivity scan above a uniform domain on sample B, revealing weak stray fields due to surface roughness \cite{supplementary}.  Dwell time per pixel is $1.5\unit{s}$ and total acquisition time is $26\unit{h}$. 
{(c)} Surface magnetization $\sigz$ reconstructed from the stray field map of panel b, given in units of Bohr magnetons ($\muB$) per $\rm{nm^2}$.  Scale bars for b and c, $2\unit{\um}$.
}
\label{fig2}
\end{figure}

Figure~\ref{fig2}(a) shows a laser-optical second-harmonic-generation (SHG) \cite{Pisarev1997} micrograph of the global domain pattern.  We observe that the domains in the bulk \CrO crystals are large, typically in the range of hundreds of micrometers, and stable below $\TN$, in agreement with earlier studies \cite{Fiebig1995}.  We find no correlation between the domain pattern and the in-plane crystal axes (Fig.~\ref{fig2}a), indicating that the domain-wall location is set by the local defect or strain distribution or is completely random.

Once the domains are localized, we acquire high-resolution magnetic imaging scans along the domain walls using NSDM microscopy (Fig.~\ref{fig2}b).  The domain wall appears as a narrow track of strong magnetic stray field in the magnetometry image; this strong field is due to the $180^\circ$ reversal of uncompensated moments near the sample surface (see Fig.~\ref{fig1}b).  Fainter features within the domains reflect residual stray fields associated with surface topography \cite{Hedrich2020}.
We do not observe any correlation between the domain wall location and the sample structure, suggesting that there are no surface-induced pinning effects.  Further, when cycling the sample through the transition temperature $\TN$, domain walls usually form in random locations of the sample with no correlation between consecutive warming-cooling cycles.

To retrieve the absolute sign of the order parameter we reconstruct \cite{supplementary} the two-dimensional surface magnetization $\sigz$ from the stray field map of Fig.~\ref{fig2}b, shown in panel c.  Here, a positive sign of $\sigz$ (dark contrast) reflects a positive \Cr surface magnetization and order parameter $\Lm$ (vice versa for $\Lp$).
We find that the correlation between SHG contrast and surface magnetization is maintained for all domain walls on all samples (Figs. S1 and S2 in Ref. \onlinecite{supplementary}).  Combined with the absence of strong magnetic features in the interior of domains, these findings directly confirm that the magnetic polarization of \CrO is robust against surface roughness \cite{He2010,Appel2019}, and that \CrO always terminates with the same \Cr surface magnetization for a given sign of the order parameter $L$.

\subsection{Domain-wall cross-section}
\begin{figure*}[t]
    \centering
    \includegraphics[width=0.65\textwidth]{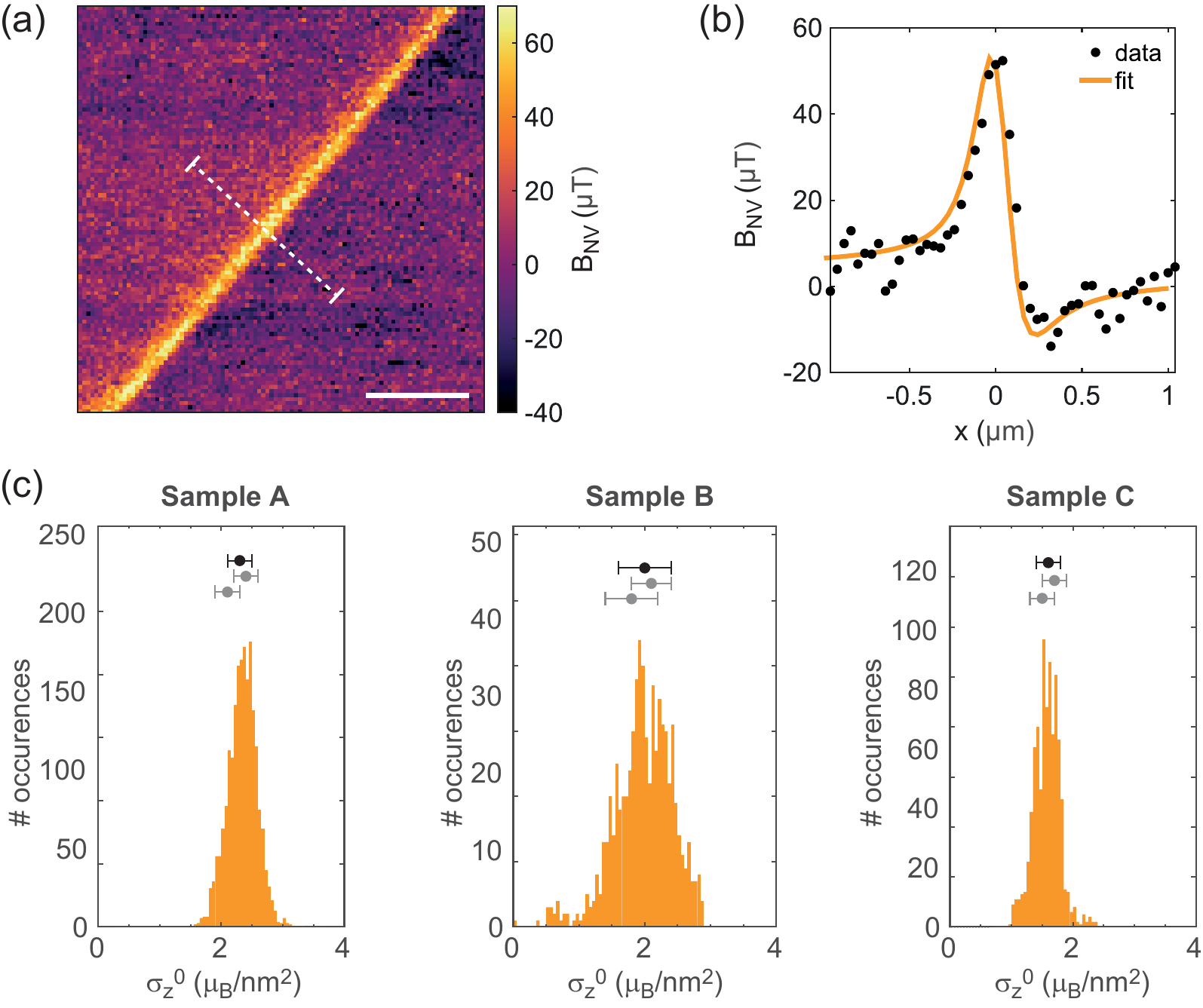}
    \caption{
		\linespread{\captionlinespread}\selectfont{}
{Quantitative measurement of domain-wall structure and surface magnetization.}
{(a)} Two-dimensional magnetometry scan of a domain wall in sample C.  Scale bar, $1\unit{\um}$.
{(b)} Cross-section along the white dashed line in panel a, showing the stray field $\BNV(x)$ as a function of the relative distance $x$ to the domain wall.  Dots are the experimental data and the solid line is a fit to the domain-wall model given by Eq.~(\ref{eq:sigwall}).  Free fit parameters are the surface magnetization $\Ms$, the domain-wall width $\Dw$, and the angle $\chi$ \cite{supplementary}.
{(c)} Histograms of $\Ms$ obtained from many line scan fits.  Mean and standard deviation (s.d.) are included above the histograms as black dots with horizontal error bars ($\pm 1\unit{s.d.}$).  Light gray bars reflect the $\Ms$ values obtained by a secondary analysis (Ref. \onlinecite{supplementary}, central bar reflects step height, and lower bar reflects integrated $\Bx$ field).  The number of line scans per histogram are 2,512 for sample A, 726 for sample B, and 1,012 for sample C.
    }
    \label{fig3}
\end{figure*}

To investigate the internal structure of a domain wall, we acquire a large number of magnetometry images along the domain wall and analyze the magnitude and spatial profile of the stray field \cite{Tetienne2015,Velez2019}.  We then compare the magnetic field along the cross section with the expected stray field from the static solution of the one-dimensional domain-wall model, Eq.~(\ref{eq:sigwall}), and compute the magnetic stray field from Eq.~(\ref{eq:sigwall}) using forward propagation\cite{supplementary}.
%
%
%
%
By fitting the computed stray field to the experimental cross-section, we obtain quantitative estimates for the surface magnetization $\Ms$, domain-wall width $\Dw$, and twist angle $\chi$.
Figure~\ref{fig3}(a,b) shows an exemplary line scan across a domain wall of sample C together with the least-squares fit. To build statistics and avoid possible cross-correlation between fit parameters, we analyze about $10^3$ line scans for each sample and validate results by a secondary data analysis (Figs. S3 and S4 in Ref. \onlinecite{supplementary}).  To exclude long-term drifts, we acquire scans along a domain wall in random order and find no temporal correlations as we proceed with the scanning.

\subsection{Surface magnetization}
Figure~\ref{fig3}c reports quantitative measurements of the surface magnetization $\Ms$.  We find a narrow distribution of $\Ms$ values ranging from $1.6(2)\unit{\muB/nm^2}$ in sample C to $2.3(2)\unit{\muB/nm^2}$ in sample A.  These values are only $15-21\%$ of the the theoretical $\Ms$ for a perfectly ordered \Cr crystal, which is $\Ms(0)=10.9\unit{\muB/nm^2}$ for the surface termination shown in Fig. \ref{fig2}(a) at zero temperature \cite{supplementary}. The low $\Ms$ is partially explained by the decay of magnetic order close to $\TN$, and is more pronounced for sample C due to the lower $\TN$.  According to Ref. \onlinecite{Appel2019}, the surface magnetization close to $\TN$ is approximately given by $\Ms(T) = \Ms(0) [1-(T/\TN)]^{0.35}$, which gives $\Ms(T)/\Ms(0)\sim 30\%$ at $T=295\unit{K}$.  Since low values for $\Ms$ have also been reported by other experimental studies \cite{Astrov1996,Appel2019}, and since we observe a narrow distribution of $\Ms$ that is uniform across the sample surface, we believe that the reduced $\Ms$ is a general and unexplained property of \CrOa.  We hypothesize that the reduced surface moment density may be due to disorder within the exposed layer of terminating \Cr ions (see Fig. \ref{fig1}(a)).

\subsection{Domain wall chirality and width}
\begin{figure}[t!]
    \centering
    \includegraphics[width=1.00\columnwidth]{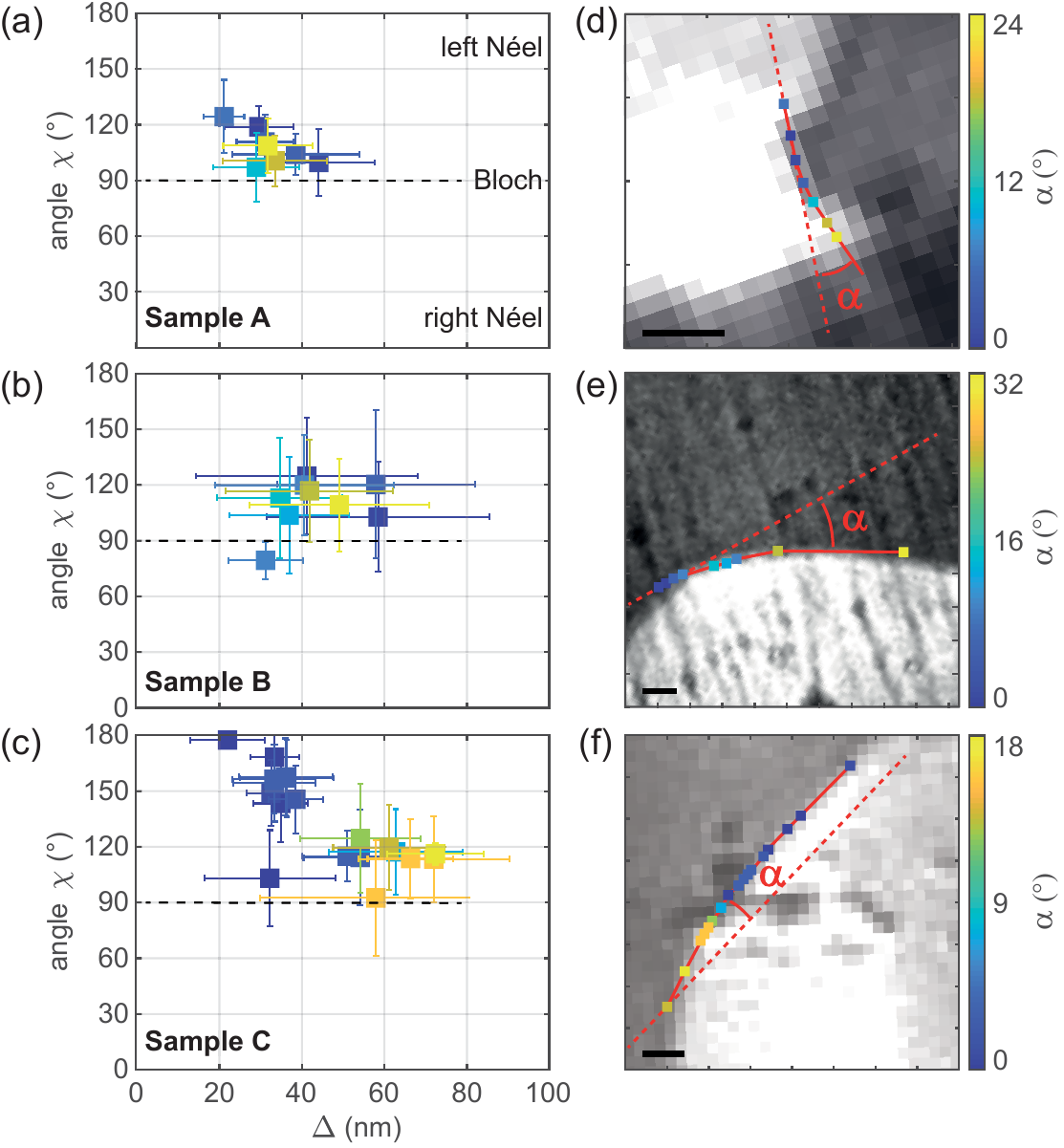}
    \caption{
		\linespread{\captionlinespread}\selectfont{}
{Observation of Bloch and N\'eel walls.}
{(a-c)} Twist angle $\chi$ plotted against the domain-wall width $\Dw$ for samples A-C.  Each point represents the data from a two-dimensional magnetometry scan.  Error bars ($\pm 1\unit{s.d.}$) are obtained by separate fits to each line of the the 2D scan and computing the standard deviation ($\text{s.d.}$) of the fit results.  Color coding reflects the propagation direction of the domain wall (see right panels).  No correlation between chirality and spatial position is evident for samples A and B, whereas a clear correlation is evident for sample C.  Mean angle and domain-wall widths are $(\chi,\Dw) = (106(6)^\circ,34(5)\unit{nm})$ for sample A, $(113(11)^\circ,45(8)\unit{nm})$ for sample B, $(143(12)^\circ,42(6)\unit{nm})$ for sample C with $\alpha<9^\circ$, and $(117(7)^\circ,65(4)\unit{nm})$ for sample C with $\alpha>9^\circ$; brackets denote standard error.
{(d-f)} SHG images of the domain-wall regions analyzed in panels a-c.  Colored squares show the scan locations.  $\alpha$ is the angle between the local propagation direction of the domain wall (red solid line) and one of the magnetic hard axes (red dashed line, see Fig. \ref{fig2}a).  Scale bars, $25\unit{\um}$.
}
    \label{fig4}
\end{figure}

Figure~\ref{fig4}a-c plots the fit results for the domain-wall width $\Dw$ and angle $\chi$ obtained from the extensive datasets recorded on samples A-C.  Each plotted $(\chi,\Dw)$ pair represents a $\sim 4\times4\unit{\um^2}$ magnetometry scan, and color-coding reflects the propagation direction of the domain wall.
For samples A and B we find all domain walls to be predominantly Bloch-like, indicated by a $\chi$ angle close to $90^\circ$ (Fig.~\ref{fig4}a,b).  The domain-wall widths are not identical, but of similar magnitude $\Dw \sim 40\unit{nm}$, and well in the range of $20-80\unit{nm}$ predicted by theory \cite{Kota2016}.  Clearly, there is no correlation between $(\chi,\Dw)$ and the spatial location or propagation direction $\alpha$ of the domain wall (see panels d,e), indicating that the crystal structure plays no role in domain-wall formation.  The consistency of the results from the two samples, which are grown independently by the same technique, confirms that our methods for quantifying the domain-wall structure are robust and reproducible.

Interestingly, sample C -- which has an unusually strong in-plane anisotropy \cite{Pisarev1997} -- shows a behavior that is distinctly different from samples A and B.  Most prominently, we find both N\'eel and Bloch walls and a pronounced dependence of the twist angle on the wall orientation.  For walls that run approximately parallel to one of magnetic hard axes (dashed lines in Fig.~\ref{fig2}a and Fig.~\ref{fig4}d-f), the domain wall has a distinct left N\'eel character (blue data points in Fig. \ref{fig4}c).  Once the angle $\alpha$ between the propagation direction and the hard axis exceeds about $9^\circ$, the wall changes to Bloch-type, and becomes similar to samples A and B.  In addition, the domain-wall width increases from $\Dw=42\unit{nm}$ in the N\'eel to $\Dw=65\unit{nm}$ in the Bloch configuration. The correlation between $(\chi,\Dw)$ and $\alpha$ is not complete, but pervasive, suggesting that a delicate balance of interactions determines the local structure of the wall.


\section{Discussion}

%
The formation of distinct Bloch and N\'eel walls in \CrO is intriguing, because in the absence of a demagnetizing field and in-plane anisotropy, the domain-wall energy of a collinear antiferromagnet is independent of the angle $\chi$ \cite{Malozemoff2016,Papanicolaou1995,Mitsumata2011,supplementary}.  Therefore, no domain-wall type is energetically favored.  In \CrOa, however, domain walls have a non-vanishing local magnetic moment associated with the spatially inhomogeneous order parameter \cite{Papanicolaou1995,Tveten2016}, giving rise to a small but non-zero demagnetizing field.  We propose that this residual demagnetizing field, which is mostly a bulk effect, is responsible for the observation of Bloch walls in samples A and B, similar to the situation encountered in uniaxial ferromagnets \cite{Hubert2008}.

The preference for Bloch walls is challenged once significant in-plane anisotropy is present (sample C).  An in-plane anisotropy favors \Cr spins aligned with the in-plane easy axis and for a sufficiently strong anisotropy, the domain wall is expected to change from Bloch to N\'eel (see Section~\ref{1Dmodel}).
Due to the three-fold crystal symmetry of \CrOa, three in-plane easy axes exist (that coincide with the crystal axes $a$, $a'$ and $b$, see Fig. \ref{fig2}a) leading to six preferred directions in $60^\circ$ intervals.  Therefore, the \Cr spins will tend to align to the nearest preferred easy direction.  The alignment is strongest when the domain wall is perpendicular to an easy axis, explaining the appearance of N\'eel walls near $\alpha\approx 0^\circ$ (blue data points in Fig. \ref{fig4}c).  Once $\alpha$ becomes larger, the in-plane anisotropy torque is reduced, and the domain wall eventually changes back to a Bloch type (yellow data points).  The critical angle where this change occurs is not well defined, but is roughly $\alpha\approx 9^\circ$.  At the same time as the domain-wall type changes from N\'eel to Bloch, the domain-wall width is expected to increase, in line with our observation (Fig. \ref{fig4}c).  Using Eq.~(\ref{eq:reductiondw}) and setting $M = s\Ms$, the one-dimensional model predicts a ratio of domain-wall widths of $\Dw^\mr{Neel}/\Dw^\mr{Bloch} = 0.85$ for \CrOa, in reasonable agreement with the experimental result of $r=0.65\pm0.10$ (Fig. \ref{fig4}c).  The good overall agreement between experiment and theory motivates the conclusion that the non-vanishing magnetic moment and in-plane anisotropy determine the domain-wall structure of \CrOa.

A final point that remains to be explained is the preference for left chiral N\'eel walls in sample C, which is also partially present in samples A and B (Fig. \ref{fig4}a-c).
Although the asymmetry is conspicuous, it is not entirely surprising given the complex magnetoelectric properties of \CrO \cite{Krichevtsov1988}.
Because the orientation of the spins in a left chiral N\'eel wall is against the stray field produced by the uncompensated magnetization of the top-most surface layers of \CrO (Fig. \ref{fig1}b), the preference for left walls cannot be attributed to a magnetostatic effect, unlike the change of a Bloch wall into a N\'eel wall observed in the near-surface region of ferromagnets \cite{Hubert2008}.  Future theoretical work shall determine whether a wall-related DMI or higher-order multispin interactions are responsible for the domain-wall chirality (see Section~\ref{1Dmodel}).  In a non-centrosymmetric environment, the DMI results in canting of the spins when $L$ has a non-zero in-plane component \cite{Dzyaloshinsky1958,Mu2019}, which -- unlike in bulk \CrO -- may be the case \textit{within} the \CrO domain wall.


\section{Summary and Outlook}

In summary, we have resolved the spin structure of $180^\circ$ domain walls in the prototype uniaxial antiferromagnet \CrOa.  We propose that the structure of the domain wall is determined by the weak energy scales provided by the non-vanishing magnetization of the wall, the in-plane magnetic anisotropy, and possibly the DMI.  Domain walls are Bloch-like in crystals with weak or negligible in-plane magnetic anisotropy, and either Bloch- or N\'eel-like in the crystal with larger in-plane anisotropy. In the latter case, the domain-wall type turns to N\'eel if the wall runs orthogonal to an in-plane easy axis, which coincides with the spin direction in the spin-flop phase of \CrO \cite{Fiebig1996}.
In agreement with simple theoretical considerations, the domain-wall width decreases from $\Dw=65\unit{nm}$ in the Bloch configuration to $\Dw=42\unit{nm}$ in the N\'eel configuration. Finally, the comparison between SHG and NSDM allows for determining the absolute sign of the order parameter in different domains, which is not possible by optical imaging alone.

Besides its fundamental interest, insight into the domain walls of collinear antiferromagnets is relevant for the development of antiferromagnetic spintronic devices that exploit current-induced domain wall motion to switch the orientation of the order parameter \cite{Shiino2016,Gomonay2016}. For example, in antiferromagnetic/heavy-metal bilayers the domain wall velocity is predicted to be zero for Bloch walls when considering only the damping-like spin-orbit torque, and nonzero but offset by a threshold current density when including the field-like spin-orbit torque \cite{Shiino2016}. In contrast, a nonzero domain-wall velocity is predicted for N\'eel walls at any current density (in the absence of pinning), making this type of wall much more efficient for achieving current-induced domain-wall displacements.  In our work, we show that the residual demagnetization field in the walls of a collinear antiferromagnet with unixial anisotropy favors the formation of Bloch walls, whereas the presence of in-plane magnetocrystalline anisotropy, likely in combination with chiral spin interactions, favors the formation of N\'eel walls. Future studies should aim at confirming the presence of a wall-related bulk DMI in \CrO and determine whether an additional interfacial DMI can be induced by proximity to a heavy metals like Pt.\\

\begin{acknowledgments}
We acknowledge T. Weber for support with the X-ray platform at the Materials Department of ETH Zurich, and N. Spaldin for insightful discussions.
This work was supported by ETH Zurich and the Swiss Competence Centre for Materials Science and Technology (CCMX).
C.D. acknowledges support by the Swiss National Science Foundation (SNFS) under grant no. 200020\_175600, by the SNFS under the NCCR QSIT, and by the European Commission under grant no. 820394 ``ASTERIQS''.
P.G. acknowledges support by the SNFS under grant no. 200020\_172775.
M.F. acknowledges support by the SNSF under grant no. 200021\_178825 and FAST, a division of the NCCR MUST ETH FAST 3 via PSP 1-003448-054.
\end{acknowledgments}

%

%







\end{document}


\begin{center}

\large

\vspace{15 mm}


\textbf{{Supplemental Material for: \\ \vspace{10 mm} ``Co-existence of Bloch and N\'eel walls in a collinear antiferromagnet''}}

\normalsize

\vspace{15 mm}

M. S. W\"ornle$^{1,2}$, P. Welter$^1$, M. Giraldo$^2$, T. Lottermoser$^2$, M. Fiebig$^2$, P. Gambardella$^2$, and C. L. Degen$^1$

\vspace{5 mm}

\textit{$^1$Department of Physics, ETH Zurich, Switzerland.}\\
\textit{$^2$Department of Materials, ETH Zurich, Switzerland.}

\end{center}



\clearpage

\section{Materials and Methods}

\subsection{Sample preparation}

We study \CrO domain walls on three bulk single crystals.  Samples A and B (Refs.~\onlinecite{Fiebig1995,Fiebig1996,Fiebig1996thesis}) are grown by the Verneuil method and oriented using a single-crystal X-ray diffractometer. The samples are cut perpendicular to the $z$-axis or (001) orientation. Subsequently, the samples are thinned down to $70\unit{\um}$. Both samples are lapped and polished, each of them following a different process. Sample A is lapped using SiC powder with $3\unit{\um}$ grain size on a cast-iron-lapping plate. Subsequently, the sample is polished following a two-step process. In the first step, the lapped surface is polished with a soft metal plate using diamond powder with $1\unit{\um}$ grain size. In the second step, a refining polishing step follows using a polyurethane polishing plate together with colloidal silicate. Here, scratches from previous mechanical treatments are removed. The sample surface is polished until it reveals a root-mean-square (rms) roughness below 1\,nm. Sample B is lapped using Al$_{2}$O$_{3}$ powder and H$_{2}$O solution. Next, the lapped surface is diamond polished until it reveals a surface with a rms roughness below 3\,nm. Sample C (Ref.~\onlinecite{Fiebig1996thesis}) is a flux-grown (001) \CrO platelet of $30\unit{\um}$ thickness. The flat as-grown surface presents a rms roughness below 0.5\,nm.  SHG images of all crystals are shown in Figs. S1 and S2.  
We create antiferromagnetic domains by cooling samples through the transition temperature $\TN$.  For samples A and B, domains are induced by magnetoelectric poling~\cite{Krichevtsov1988}.  In sample C, different domain patterns spontaneously form when the sample is cooled through $\TN$.

\subsection{Second harmonic generation (SHG) measurements}

SHG microscopy exploits an interference contrast of frequency-doubled optical photons in domains of opposite magnetic polarization to reveal the domain pattern~\cite{Fiebig1995}.  A magnetic contribution to the frequency-doubled light wave coupling linearly to the antiferromagnetic order parameter $\pm L$ interferes with a frequency-doubled crystallographic background contribution which identifies the two antiferromagnetic domain states by their different brightness~\cite{Fiebig1994}.  We use a transmission SHG setup to acquire the SHG images, in which we use a Coherent Elite Duo laser system, which emits 120\,fs pulses at a repetition rate of 1\,kHz.  An optical parametric amplifier tunes the wavelength to excite the bulk \CrO samples with a photon energy of 1.033\,eV and a pulsed energy of 80\,$\mu$J. The crystals are excited in transmission and at normal incidence by an unfocused circularly-polarized laser beam. Right-handed circularly-polarized light denote the clockwise rotation of the electric-field vector of light with respect to its propagation direction. The opposite follows for left-handed circular polarization. A camera lens is used to collect the SHG signal. Optical filters are added to select the SHG spectral wavelength, suppressing the fundamental beam and higher-harmonic contributions. SHG light is detected at room temperature with a Jobin-Yvon, back-illuminated, deep-depletion digital camera with a near-infrared detector chip of 1024$\times$256 pixels. The camera is cooled with liquid nitrogen to reduce thermal noise.

\subsection{Nanoscale scanning diamond magnetometry (NSDM) measurements}

Scanning NV magnetometry measurements are carried out on a user-facility instrument, built in-house~\cite{ccmx}, and under ambient conditions.  The instrument uses $520\unit{nm}$ laser light and $2.76\unit{GHz}$ to $2.91\unit{GHz}$ microwave pulses to detect the NV center spin resonances.  Laser illumination is kept below $20\unit{\uW}$ to avoid laser-induced heating of the sample.  The spin resonance frequency is determined by sweeping the microwave frequency and fitting a Lorentzian function to the optically-detected magnetic resonance spectrum.  Four different diamond probes (QZabre LLC) of $\sim 22\%$ CW ODMR contrast at a measurement count rate of $\sim 200\unit{kC/s}$ are used.  The sensitivity of these probes (as determined from the average least-squares variance of the center frequency) is $1.7\unit{\uT}$ for an integration time of 6.4 seconds per pixel.  All scans are performed on the \CrO surface pointing towards the camera in the SHG experiment.

We use both continuous and pulsed ODMR protocols~\cite{Dreau2011,Schirhagl2014} on either transition ($m_S=0$ to $m_S=\pm1$) of the NV center. A small external bias field of $\sim 4\unit{mT}$ is applied to split the spin resonances; this small bias field is not expected to influence the \CrO physics.  To convert the measured spin resonance frequency $f$ to units of magnetic field, we compute
\begin{align}
\BNV = \frac{f_0 - f}{28.02\unit{MHz/mT}}
\end{align}
where $f_0$ is the mean frequency taken over the entire scan, which is approximately the frequency far from the sample surface.  We recall that NV magnetometry provides one vector component of the magnetic field, $\BNV = \vece\cdot\vecB$, which is the projection of $\vecB$ onto the anisotropy axis $\vece = (\ex,\ey,\ez)$ of the spin.
The unit vector $\vece = (\sin\thNV\cos\phiNV,\sin\thNV\sin\phiNV,\cos\thNV)$ corresponds to the symmetry axis (N-V axis) of the NV center, as expressed by the laboratory frame angles $\thNV$ and $\phiNV$.  The sensor vector orientation is pre-determined for each tip using an external field sweep.  The stand-off distance $z$ between NV center and the sample surface is measured by independent calibration scans over a magnetized Co stripe before and after the \CrO scans~\cite{Tetienne2015,Velez2019}.  For our probes, ($\thNV,\phiNV,z$) is ($55^\circ$, $270^\circ$, $73\pm7\unit{nm}$) for tip A,   ($55^\circ$, $180^\circ$, $64\pm4\unit{nm}$) for tip B, ($55^\circ$, $176^\circ$, $65\pm3\unit{nm}$) for tip C and ($55^\circ$, $176^\circ$, $68\pm8\unit{nm}$) for tip D.

\section{Data analysis}

\subsection{Effective surface magnetization $\Ms$}

Antiferromagnetic order in the form of vertically alternating layers of oppositely polarized ions leads to an effective surface layer magnetization on the top and bottom surfaces of the crystal, in analogy to the bound surface charge appearing for a polarized dielectric~\cite{Belashchenko2010,Andreev1996,He2010}.  To calculate the surface magnetization, we assign the alternating layers of opposite polarization to two oppositely magnetized volumes, each with magnetization $M_s = n m/V$, vertically shifted with respect to each other by $s$.  Here, $n$ is the number of ions per unit cell and polarization direction, $m$ is the magnetic moment per ion, and $V$ is the volume of the unit cell.  Within the bulk, the magnetization of the two volumes is exactly compensated, except in two thin layers of thickness $s$ at the top and bottom of the body.  Thus, the bulk antiferromagnetic order appears like an magnetized surface layer at the top and bottom of the crystal, with an effective layer magnetization of
\begin{align}
\Ms = d M_s = \frac{n m s}{V} \ .
\end{align}
For thick crystals a local magnetic probe only detects the stray field of the top layer.

\CrO has a hexagonal unit cell with a lattice constant of $a_\mr{hex}=4.96\unit{\AA}$, corresponding to a side length of $a/\sqrt{3}=2.86\unit{\AA}$, a height of $c=13.6\unit{\AA}$, a hexagonal surface area of $A=\frac{3\sqrt{3}}{2}a^2=21.3\unit{\AA^2}$, and a volume of $V=Ac=290\unit{\AA^3}$ (Refs. \onlinecite{Fiebig1996thesis,He2010}).
The hexagonal unit cell is constructed from six vertically stacked \O planes.  Each \O plane has two nearest \Cr ions of opposite magnetic polarization located $0.941\unit{\AA}$ above or below the plane, respectively, therefore $s=1.88\unit{\AA}$ (see Fig.~2a and Ref.~\onlinecite{Fiebig1996thesis}).  Accounting for the 12 \Cr ions per unit cell, $n=6$ for each orientation.  Assuming a moment of $m=2.8\unit{\muB}$ per \Cr ion~\cite{Madelung2000}, we calculate a surface magnetization of
\begin{align}
\Ms = \frac{6 \times 2.8\unit{\muB} \times 0.188\unit{nm}}{0.290\unit{nm^3}} = 10.9\unit{\muBnm} \ .
\end{align}
This is slightly less than what one would expect from one monolayer of \Cr ions, which would have a magnetization of $m/A=12.1\unit{\muBnm}$.

\subsection{Transformations between surface magnetization and magnetic field}

Using the relations between magnetization and magnetic stray field for two-dimensional thin films~\cite{Beardsley1989}, we can reconstruct the surface magnetization $\sigz(x,y)$ and vector magnetic field $\vecB(x,y)$ from the measured stray field component $\BNV(x,y)$.  We perform transformations in Fourier space.  The magnetic vector field $\vecB$ associated with the magnetization $\vecsig$ is given by
\begin{align}
\rowvec{\hatBx,\hatBy,\hatBz} = \frac12 \mu_0 e^{-kz} \rowvec{-\kx\hatsigk-i\kx\hatsigz, -\ky\hatsigk-i\ky\hatsigz, -ik\hatsigk+k\hatsigz}
\label{eq:BfromM}
\end{align}
where $\kx$, $\ky$ are the in-plane $k$-vectors, $k=(\kx^2+\ky^2)^{1/2}$, $\hatsigk = (\kx\hatsigx+\ky\hatsigy)/k$, and hat symbols denote Fourier transforms in $x$ and $y$. $z$ is the stand-off distance of the sensor and $\mu_0=4\pi\ee{-7}\unit{Tm/A}$.  For a line scan in $x$ direction, scanned across a domain wall extending in $y$ direction, the magnetic field is
\begin{align}
\rowvec{\hatBx,\hatBy,\hatBz} = \frac12 \mu_0 e^{-kz} \rowvec{ -\kx\hatsigx-i\kx\hatsigz, 0, -ik\hatsigx + k\hatsigz } \ ,
\label{eq:BfromMline}
\end{align}
where now $k = |\kx|$.
Likewise, we can recover the magnetic vector field $\vecB$ from the measured projection $\BNV$ as
\begin{align}
\rowvec{\hatBx,\hatBy,\hatBz} = \frac{1}{\kNV} \rowvec{ i\kx, i\ky, -k } \hatBNV
\label{eq:vecBfromB}
\end{align}
where $\kNV = (i\ex\kx+i\ey\ky-\ez k)$ and $(\ex,\ey,\ez)$ is the vector orientation of the sensor.  Finally, under the assumption that the magnetization is fully out-of-plane ($\sigx=\sigy=0$), we can reconstruct $\sigz$ from the stray field $\BNV$,
\begin{align}
\hatsigz = - \frac{2W \hatBNV}{\mu_0 e^{-kz}\kNV}
\label{eq:MfromB}
\end{align}
where $W=W(k)$ is a suitable window function (here a Hann function) that provides a high-frequency cutoff.
Although our \CrO films do have an in-plane component in the vicinity of the domain wall, the reconstructed $\sigz$ still accurately reproduces the domain pattern and surface magnetization $\Ms$.

\subsection{Magnetic field from surface roughness}

Surface roughness leads to tiny stray fields at topographic steps, as sketched in Fig.~1b.  The magnetic field produced at a step of height $h$ corresponds to the differential field of two magnetized layers located at $z$ and $z+h$.  According to Eq.~\ref{eq:BfromMline}, the $\Bz$ field of the step is  given by
\begin{align}
\hatBz = \frac12 \mu_0 e^{-kz} k h (k\hatsigz) \ .
\end{align}
For a simple order-of-magnitude estimate of the stray field, we look at the Fourier component of $\sigz$ that produces the strongest $\Bz$.  This occurs for $k = 2/z$.  For this Fourier component, the amplitude of $\Bz$ is
\begin{align}
\Bz = \frac{\mu_0 h 2e^{-2}\Ms}{z^2} \approx \frac{0.2707 \mu_0 h\Ms}{z^2}
\end{align}
For our \CrO crystals, where $\Ms\approx 2\unit{\muBnm}$, and using $z=68\unit{nm}$, we find $\Bz/h \approx 1.4\unit{\uT/nm}$.  For an rms surface roughness of $3\unit{nm\rms}$ we therefore expect stray field fluctuations of $\sim 5\unit{\uT}$, in good agreement with the experimental $7\unit{\uT\rms}$ (Fig.~2d).

\subsection{Fitting of line scans}

We model the domain wall as presented in Eq. 1 in the main text. The stray field is then computed via Eq.\ \ref{eq:BfromMline}. The resulting model features 7 parameters: the effective surface magnetization $\Ms$, the position of the domain wall $\xo$, its width parameter $\Dw$ and twist angle $\chi$, and the sensor geometry $(z, \thNV, \phiNV)$. Since $z$, $\thNV$, $\phiNV$ have been determined separately at this point, they are left fixed in the following least-squares optimization, leaving only $\sigz$, $\xo$, $\Dw$ and $\chi$ as free parameters.

The initial value of $\Ms$ is determined by estimating the surface magnetization using the two complementary methods (step height, integration of $\Bx$) described below. The initial value for width and chirality are set to $\Dw = 40 \unit{nm}$ and $\chi = 90^\circ$.  We checked that other starting values did not significantly alter the fit results.  The fitting procedure is repeated for each individual line scan.

\subsection{Complementary methods for estimating $\Ms$}

We use two complementary methods for estimating the \CrO surface magnetization $\Ms$ from a stray field scans across domain walls:

\textit{Step height in reconstructed $\sigz$ map:} We reconstruct the surface magnetization $\sigz(s)$ using Eqs.~(\ref{eq:BfromMline}) and (\ref{eq:MfromB}).  The step height at the domain wall is $2\Ms$.

\textit{Integration of $\Bx$:} We assume a domain wall extending along the $y$ direction.  We compute the $\Bx(x)$ component of the stray field from $\BNV(x)$, using the known orientation of the sensor $(\thNV,\phiNV)$ and Eq.~(\ref{eq:vecBfromB}).  The integrated $\Bx(x)$ is then equal to $\mu_0\Ms$, irrespective of the stand-off $z$ and the domain-wall profile and chirality.  To explain this, assume an out-of-plane magnetized film with magnetization $\vecsig(x')$ and a domain wall centered at $x=0$ and extending along the $y$-direction.  The step edge can have a $\sigx$ or $\sigy$ component.  The magnetic field $\Bx$ produced by the magnetization element $\mathrm{d}x' \vecsig(x')$ is
\begin{align}
\mathrm{d}B(x) = \frac{\mu_0 j_y(x') t \mathrm{d} x' z}{2\pi[(x-x')^2+z^2]}
\end{align}
where $j_y(x') t = [\vec\nabla\times\vecsig]_y(x') = -[\partial_x \sigz](x')$ is the bound current element associated with $\vecsig(x')$ and $t$ is the film thickness ($t\ll z$).  The total magnetic field at position $x$ is
\begin{align}
B(x)
= \int_{-\infty}^{\infty} \mathrm{d}x' \frac{\mu_0 j_y(x') t z}{2\pi[(x-x')^2+z^2]}
= -\int_{-\infty}^{\infty} \mathrm{d}x' \frac{\mu_0 z}{2\pi[(x-x')^2+z^2]} \, [\partial_x \sigz](x')
\end{align}
and the integrated $B(x)$ is
\begin{align}
\int_{-\infty}^{\infty} \mathrm{d}x B(x)
&= - \left( \int_{-\infty}^{\infty} \mathrm{d}x'' \frac{\mu_0 z}{2\pi[(x'')^2+z^2]} \right)
\, \left( \int_{-\infty}^{\infty} \mathrm{d}x' [\partial_x \sigz](x') \right)  \\
&= -\frac{\mu_0}{2} \, [\sigz(+\infty)-\sigz(-\infty)]
= \mu_0 \Ms
\end{align}
where we have used Fubini's theorem and the last equation is for a domain wall where $[\sigz(+\infty)-\sigz(-\infty)] = -2\Ms$.  

\subsection{Complementary method for estimating $\Dw$ and $\chi$}

For a fixed pair ($\chi$,$\Dw$), we only fit $\xo$ to the data, and record the residual sum of squares (RSS). The surface magnetization is determined for each line scan by the previously introduced three complementary methods. The RSS is a measure of the likelihood. Indeed, assuming Gaussian errors, the log-likelihood is given by
\begin{align}
	\ln\mathcal{L} = \ln\left(\frac{1}{2\pi\sigma^2}\right)\frac{n}{2} - \frac{1}{2\sigma^2}\mathrm{RSS}
\end{align}
Here, $\sigma$ is the standard deviation describing the error of a single data point, and $n$ is the number of data points. We can compare the relative likelihood of two models 1 and 2 (\ie two pairs of $\Dw$ and $\chi$) by estimating $\sigma^2_i = \mathrm{RSS}_i/n$, $i\in\{1,2\}$, giving
\begin{align}
	\ln\mathcal{L}_1 - \ln\mathcal{L}_2 = -\frac{n}{2}\ln\frac{\mathrm{RSS}_1}{\mathrm{RSS}_2}
	\label{eq:relativeLL}
\end{align}
We choose model 2 as the best model (i.e. the least squares solution), so that Eq.\ \ref{eq:relativeLL} is normalized to 0. To consider the data from all scans, we sum the RSS of each line and scan, and set $n$ to be the total number of data points.

\clearpage
\section{Supplementary Figures}

%
\begin{figure}[h!]
\includegraphics[width=0.8\textwidth]{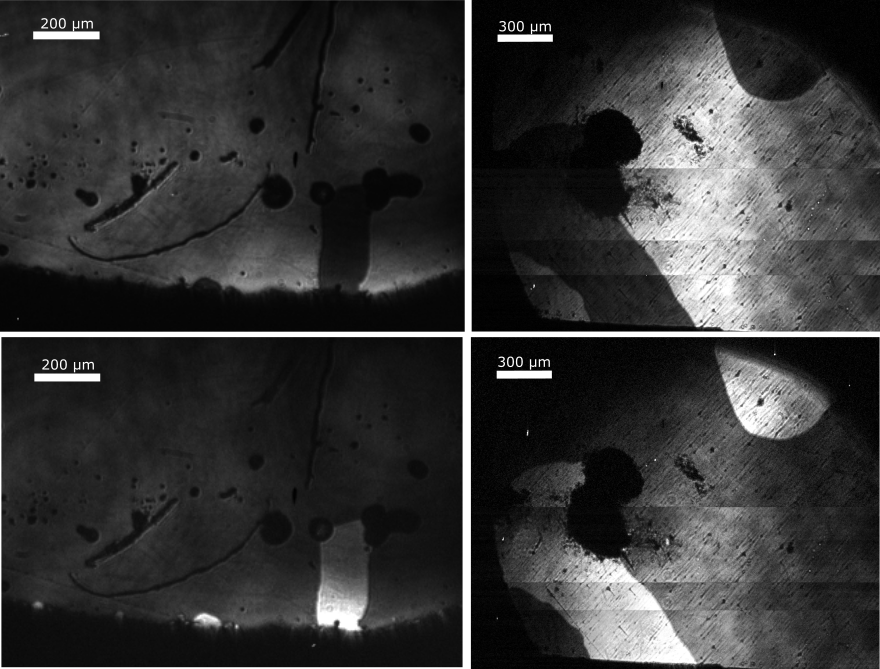}
\caption{SHG microscopy images of sample A (left panels) and sample B (right panels).  Upper panels used left-handed circular polarization, lower panels used right-handed circular polarization.  
\label{fig:figS1}
}
\end{figure}

\clearpage
%
\begin{figure}[h!]
\includegraphics[width=0.8\textwidth]{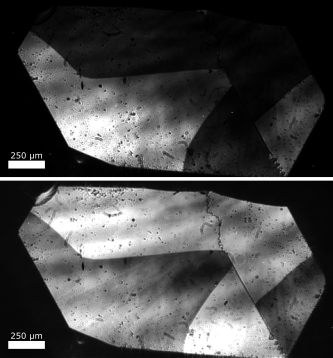}
\caption{SHG microscopy images of sample C.  Upper panel used left-handed circular polarization, lower panel used right-handed circular polarization.
\label{fig:figS2}
}
\end{figure}

\clearpage
%
\begin{figure}[h!]
\includegraphics[width=0.8\textwidth]{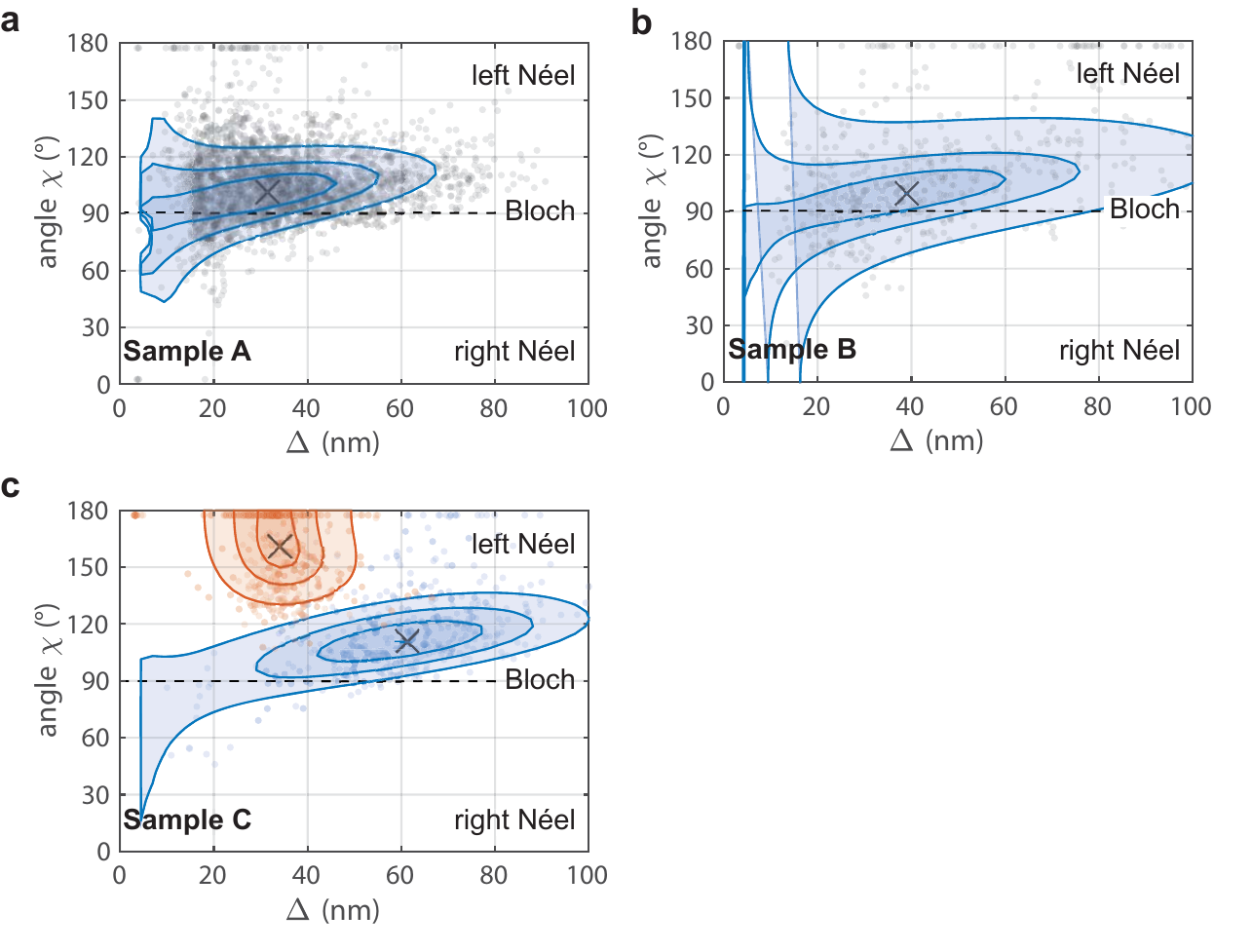}
\caption{Maximum likelihood estimates for domain wall width $\Dw$ and twist angle $\chi$, as explained in the Methods section.  Gray dots are fit results from individual line scans.  Colored contours are maximum likelihood isolines containing 75\%, 50\%, and 25\% of datapoints.  The most likely ($\chi,\Dw$) pair is indicated by a central cross. For sample C, datasets with $\alpha>9^\circ$ (blue) and $\alpha<9^\circ$ (red) are analyzed separately.
\label{fig:figS3}
}
\end{figure}

\clearpage
%
\begin{figure}[h!]
\includegraphics[width=0.8\textwidth]{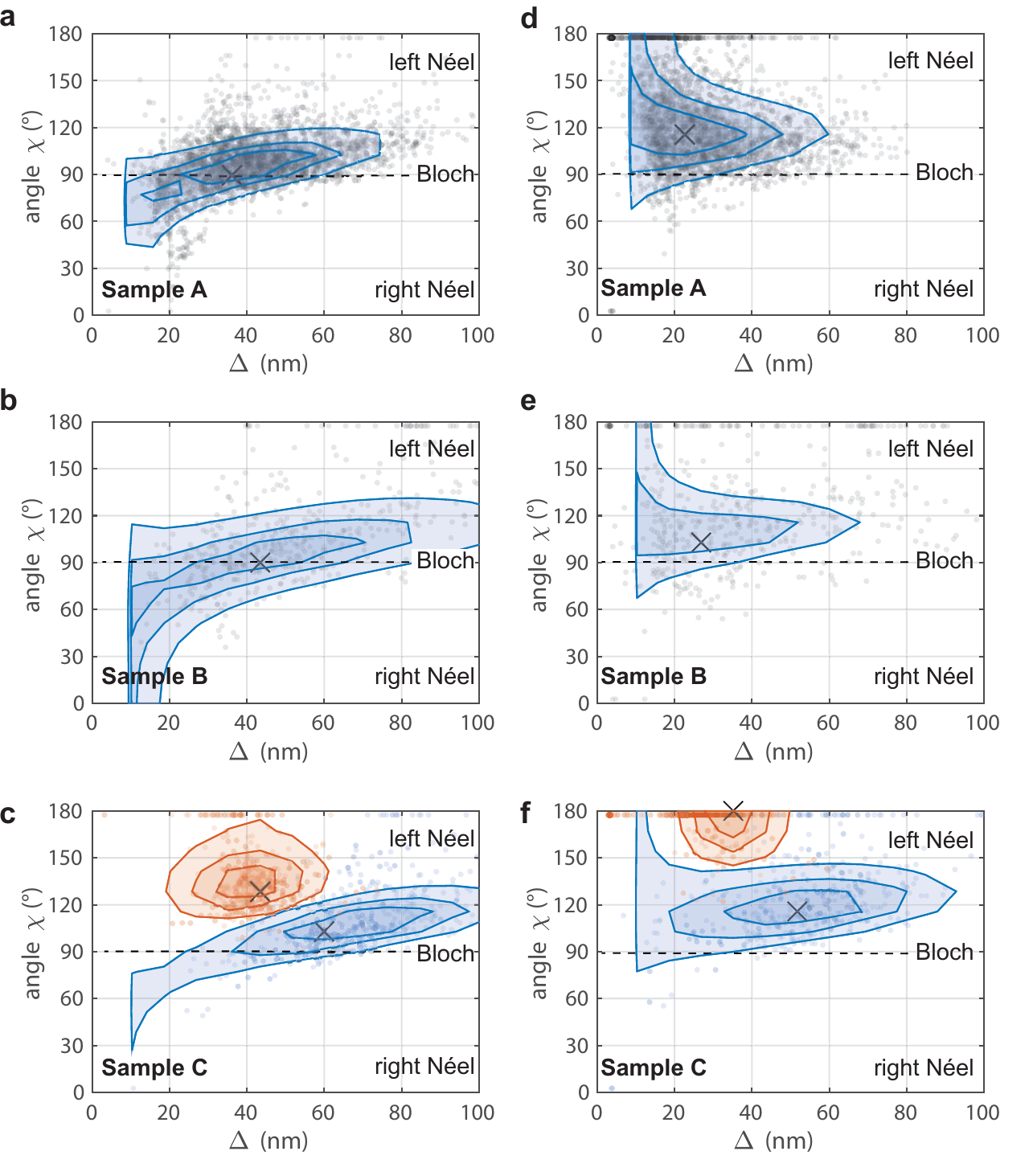}
\caption{Maximum likelihood estimates for domain wall width $\Dw$ and twist angle $\chi$ for upper and lower bound stand-off distances $z\pm 10\unit{nm}$, where $z$ is the calibrated stand-off distance.  
{\bf a-c} Maximum likelihood estimates for the lower bound $z-10\unit{nm}$.
{\bf d-f} Maximum likelihood estimates for the upper bound $z+10\unit{nm}$.
Data points, contours and central cross are as with Fig. S3.
We note that the our observation -- the presence of Bloch-like walls in samples A and B, and mixed Bloch and N\'eel walls in sample C -- is valid within the uncertainty of the sample-sensor distance.  The $z\pm 10\unit{nm}$ bounds are a conservative estimate, as all probes showed a calibration error of $\leq 8\unit{nm}$ (see Methods). 
\label{fig:figS4}
}
\end{figure}

\clearpage

%